# Cross-Disciplinary Perspectives on Youth Digital Well-Being Research: Identifying Notable Developments, Persistent Gaps, and Future Directions

Katie Davis,[a] Morgan Anderson,[a] Chia-chen Yang,[b] Sophia Choukas-Bradley,[c] Beth T. Bell,[d] Petr Slovak[e]

[a]University of Washington, Seattle, WA, USA, [b]Oklahoma State University, OK, USA, [c]University of Pittsburgh, Pittsburgh, PA, USA, [d]University of York, York, UK, [e]King's College London, London, UK

Corresponding author:
Katie Davis
Box 352840
Mary Gates Hall, Suite 370
University of Washington
Seattle, WA 98195
USA

kdavis78@uw.edu

ORCID: https://orcid.org/0000-0001-8794-8651

This work was supported in part by a grant from the Global Innovation Fund at the University of Washington.

The authors wish to thank the Global Innovation Fund at the University of Washington for supporting this work, as well as the researchers who participated in the interviews and convening that formed the basis for this paper.

Author bios:

**Dr. Katie Davis** is Associate Professor at the University of Washington (UW) Information School, Adjunct Associate Professor in the UW College of Education, and a founding member and Co-Director of the UW Digital Youth Lab. Davis investigates the impact of digital technologies on young people's learning, development, and well-being, and co-designs positive technology experiences for youth and their families. She holds two master's degrees and a doctorate in Human Development and Education from Harvard Graduate School of Education.

**Dr. Morgan Anderson** is a Postdoctoral Clinical Fellow for the Evidence Based Treatment Centers of Seattle (EBTCS). Her research investigates how to leverage technology to identify and serve students' social emotional needs in a more proactive, equitable way. Dr. Anderson holds a PhD in School Psychology from the University of Washington.

**Dr. Chia-chen Yang** is an Associate Professor of Educational Psychology and Director of the Communication Technologies and Youth Development Lab at Oklahoma State University. Her research focuses on the psychosocial development of young people in the digital age, examining how adolescents and emerging adults use digital media and how this usage is associated with their identity development, social relationships, and socio-emotional well-being. She earned her Ph.D. in Educational Psychology from the University of Wisconsin-Madison.

**Dr. Sophia Choukas-Bradley** is a licensed clinical psychologist, Director of the Teen and Young Adult Lab, and Assistant Professor of Psychology in the Developmental, Clinical, and Social Psychology programs at the University of Pittsburg. Dr. Choukas-Bradley also serves as the Clinical Psychology Lead for the Center for Digital Thriving at Harvard Graduate School of Education.

**Dr. Beth Bell** is a Lecturer in Mental Health and Wellbeing at the University of York. Her research adopts a mixed-methods interdisciplinary lens to understanding the risks and opportunities afforded by digital technologies in relation to youth wellbeing and mental health. She holds a PhD in Psychology from the University of Sussex.

**Dr. Petr Slovak** is a Senior Lecturer in Human-Computer Interaction (HCI) at King's College London, affiliated with the Informatics department. Additionally, he holds visiting roles within the Child and Adolescent Psychology at KCL and the Experimental Psychology & Human-centered AI groups at the University of Oxford. His research primarily concerns the design and evaluation of digital mental health interventions. Spanning both preventative and clinical domains, he has contributed insights into how socio-technical design can influence the development of skills in daily life, particularly for parents, children, and youth.

## Abstract

**Aims.** The purpose of this paper is to provide a broad, multi-disciplinary overview of key insights, persistent gaps, and future paths in youth digital well-being research from the perspectives of researchers who are conducting this work.

**Demographics and Settings.** Participants included 38 researchers representing diverse disciplinary fields from universities and research institutes spanning 12 countries.

**Methodology.** We conducted semi-structured interviews with 25 researchers via Zoom and hosted a two-day virtual convening with 26 researchers, 13 of whom had participated in the interviews.

**Analysis.** We employed reflexive thematic analysis to analyze the interview and convening data.

**Findings.** Our analysis surfaced three notable developments in youth digital well-being research: (1) greater nuance in measures of young people's social media experiences, (2) increased representation of diverse adolescents' experiences, and (3) a growing number of interventions targeting youth digital well-being. Persistent gaps include a relative lack of cross-cultural work, particularly in the Global South, as well as culturally sensitive interventions that fit the needs and contexts of diverse youth.

**Implications.** By examining existing research through the perspectives of researchers spanning multiple fields, we contribute a cross-disciplinary view on the current state of knowledge and identify priorities for youth digital well-being research.

**Introduction**

Over the last fifteen years, social media and smartphones have secured their place as central elements in adolescents' lives. Concurrently, research examining their impact on youth well-being has reached the status of a well-investigated topic. The review articles published in recent years (e.g., Choukas-Bradley et al., 2023; Haidt & Twenge, ongoing; Hamilton, Nesi & Choukas-Bradley, 2022; Jensen, George, Russell & Odgers., 2019; Keles, McCrae & Grealish., 2020; Nesi, 2020; Odgers & Jensen, 2020; Odgers, Schueller & Ito, 2020; Orben, 2020; Sala, Porcaro, & Gomez, 2024; Valkenburg, Beyens, Meier, & Vanden Abeele, 2022) as well as reports such as the American Psychological Association's recent health advisory on social media in adolescence (McCabe et al., 2023) point to a collective recognition that it is time to take stock of what we know about technology's impact on youth well-being and chart a path forward. We contribute to this synthesis and path-charting work by investigating the perspectives of researchers across disciplines to distill key insights from existing research and identify persistent challenges, gaps, and open questions.

The current landscape of youth digital well-being research encompasses a variety of disciplines, methodologies, and definitions of *youth*, *digital*, and *well-being*. Existing review articles indicate a dominant focus on examining relationships between adolescents' social media use and mental health outcomes such as depression and anxiety (e.g., Choukas-Bradley et al., 2023; Haidt & Twenge, ongoing; Jensen et al., 2019; Nesi, 2020; Odgers & Jensen, 2020; Odgers et al., 2020; Orben, 2020; Sala et al., 2024; Valkenburg et al., 2022). In the current work, we use *well-being* to refer both to the absence of psychopathology (e.g., depression, anxiety) and the presence of positive indicators such as positive affect and life satisfaction (Yang et al., 2021). This view is consistent with the World Health Organization's definition of mental health, which includes both the absence of mental disorders and "a state of well-being in which every individual realizes his or her own potential, can cope with the normal stresses of life, can work productively and fruitfully, and is able to make a contribution to her or his community" (World Health Organization, 2005, p.2, as quoted in Meier & Reinecke, 2021). Thus, we sometimes use well-being and mental health interchangeably depending on the language used by participants and cited literature. We use the term *youth digital well-being* to characterize well-being in the context of adolescents' (13-18 years) networked communications, especially their social media use.

We conducted in-depth interviews with 25 researchers whose work addresses some aspect of youth, technology, and well-being. We then hosted a two-day virtual convening with 26 researchers, 13 of whom had participated in the interviews. Participating researchers came from universities spanning 12 countries and multiple disciplines, including developmental science, clinical psychology, communication, human-computer interaction, and education, among others. In both the interviews and the convening, we invited researchers to reflect on the current knowledge base in their respective fields and where they think research should go next.

Our conversations in both the interviews and two-day convening centered on the mental health outcomes associated with adolescents' social media use, such as anxiety and depression. Within this dimension of youth digital well-being, we identified three focal topics related to the current state of research related to youth digital well-being: *measurement*, *representation*, and *intervention*. For each topic, researchers discussed notable developments in their respective fields, pointed to persistent gaps, and offered generative ideas for future research directions.

By examining existing research through the perspectives of researchers spanning multiple fields, we contribute a cross-disciplinary view on the current state of knowledge and identify future priorities for youth digital well-being research.

**Twenty Years of Researching Teens and Social Media**

Research on adolescents' social media experiences follows the rise of social media platforms (as we know them today) in the early-to-mid-2000s. In 2007, approximately 93% of U.S. teens were online and 55% of those online teens used social media sites such as MySpace and Facebook (Lenhart et al., 2007). By 2023, researchers were no longer asking U.S. teens *whether* they used social media but instead *which* platforms they used and how often. For instance, about half of U.S. teens reported using Instagram and Snapchat daily in 2023, and even more (58%) said they used TikTok everyday (Pew Research Center, 2023).

During the early years of social media, research focused on teens' use of MySpace and Facebook to craft online identities, connect with friends, and participate in communities of like-minded people (e.g., boyd, 2007, 2014; Davis, 2010, 2012a, 2012b; Gardner & Davis, 2013; Ito et al., 2099). Researchers explored developmentally salient processes such as the relationship between teens' online and offline identities (e.g., Davis, 2014), self-disclosure through computer-mediated communication (e.g., Schouten et al., 2007), and how youth with marginalized identities find support and acceptance in online communities (e.g., boyd, 2014). From the beginning, researchers have shown interest in understanding the relationship between youth's social media experiences and mental well-being. This interest is evident in the proliferation of published articles exploring the mental health effects of cyberbullying (e.g., Giumetti & Kowalski, 2022), body image concerns (e.g., Choukas-Bradley et al., 2020; Maheux et al., 2022a), digital stress (e.g., Nick et al., 2022), and problematic social media use (e.g., Tomczyk & Hoferichter, 2022).

As social media platforms and the teens who use them have changed, so too has the research landscape. For instance, as social media platforms have attracted more users across the globe, researchers have broadened their focus to include youth from different cultures and countries (e.g., Ghai et al., 2022; Magis-Weinberg et al., 2021a, 2021b). Researchers have also broadened their focus when it comes to the dimensions of social media use they investigate and the methods they use. For instance, as more researchers in the field of human-computer interaction have turned their attention to social media, there has been an increased focus on the role of design in shaping social media experiences (e.g., Baughan et al., 2022; Charmaraman & Delcourt, 2021; Davis et al., 2023). This attention to technology design has helped provide a more nuanced understanding of the interaction between teens and the designed environments of different social media platforms. With respect to methods, the shift towards person-specific analyses (e.g., Jensen et al., 2019; Pitt et al., 2021; Valkenburg et al., 2021; Verbeij et al., 2022) has also contributed to a more nuanced understanding of individual variation in teens' social media experiences.

Amidst the growth of social media platforms and research focused on them, public concern about social media's impact on youth has also risen, adding a sense of urgency to research endeavors. In 2023, the U.S. Surgeon General issued a public health advisory warning of the potential for social media platforms to harm adolescents' mental health (Office of the Surgeon General, 2023). Laws such as the UK's Age-Appropriate Design Code and the proposed Kids Online Safety Act in the U.S. aim to compel tech companies to take steps to safeguard young people's safety and well-being. In the United States, a coalition of 42 states Attorneys

General have filed a lawsuit against Meta, the parent company of Instagram and Facebook, for alleged harm to young users (Stiffler, 2023). These developments underscore the need for research that can provide insight into social media's effects on young people, as well as clear direction on how to support positive social media experiences.

**Social Media's Effects on Adolescent Well-Being**

Over the last twenty years, researchers have gained a better understanding of the benefits and risks associated with adolescents' social media use, particularly in the context of developmental milestones such as identity development and peer relationships. Benefits include opportunities for social connection (e.g., Nesi et al., 2018), a relatively safe space for identity expression and exploration (e.g., Granic et al., 2020), the ability to pursue interests and develop new skills (e.g., Ito et al., 2009), and opportunities for civic engagement (e.g., Middaugh et al., 2017). Marginalized youth may benefit in distinct ways. For instance, adolescents with LGBTQ+ identities who live in rural areas use social media to find and become part of online communities that include other LGBTQ+ youth (Berger et al., 2021; Karim et al., 2022). Within these communities, adolescents experience both social connection and affirmation of their identities, something that may be difficult to experience in their offline contexts (Paceley et al., 2022; Selkie et al., 2020).

Researchers have also documented a variety of risks associated with adolescents' social media use, including cyberbullying (e.g., Giumetti & Kowalski, 2022), digital stress (e.g., Nick et al., 2022; Steele et al., 2020), body image concerns (e.g., Choukas-Bradley et al., 2020; Maheux et al., 2022a), privacy breaches (e.g., Cho, 2018), and exposure to hate speech (e.g., Costello & Hawdon, 2020). These risks are not evenly distributed among adolescents. Girls who tend to engage in social comparison are more vulnerable to body image concerns when exposed to manipulated photos on social media (Kleemans et al., 2018). Girls are also more likely to experience negative mental health outcomes related to digital stress (Nesi & Prinstein, 2015). Youth with marginalized identities also face distinct risks. For instance, LGBTQ+ youth are at particular risk of cyberbullying victimization (Mkhize et al., 2020), and Black youth are vulnerable to exposure to violent, racist content and even direct personal harassment, placing them at increased psychological risk (Costello & Hawdon, 2020; Tanksley, 2019; Tynes et al., 2020).

**Theoretical and Methodological Contributions to Understanding Teen Social Media Use**

The research findings described above point to the value of examining social media within the broader context of adolescents' lives, relationships, and cultural communities (Davis, 2023). The neo-ecological theory proposed by Navarro and Tudge (2022) provides a useful theoretical framework for understanding the person- and context-specific quality of teens' social media experiences. The neo-ecological theory extends Bronfenbrenner's original bioecological theory by considering the impact of technology and virtual environments on adolescent development.

Navarro and Tudge (2022) identify two kinds of microsystems, one virtual and one physical, and discuss how they interrelate and influence each other. Notably, individuals can inhabit more than one microsystem simultaneously. Thus, adolescents' social media experiences will differ in part according to differences in the offline microsystems surrounding their social media use. Within the broader context of the macrosystem, neo-ecological theory describes how adolescents' use of online platforms is deeply intertwined with cultural and societal norms, shaping the values, motivations, and goals they bring to their social media experiences. Person-level characteristics provide additional sources of variation in adolescents' social media experiences. These include adolescents' behavioral dispositions (e.g., curiosity, impulsiveness), the presence and absence of personal resources (e.g., knowledge, skills, illness), and observable features such as gender, age, skin color, and attractiveness.

This theoretical view of adolescents' social media use and its impact on development underscores the need to employ methods that capture various dimensions of young people's social media experiences. Qualitative research has attended to such nuances from the beginning (e.g., boyd, 2014; Ito et al., 2009), but quantitative research has tended to focus on aggregate-level trends that do not distinguish among individual adolescents and platforms. However, quantitative researchers are increasingly turning to more person-specific studies that capture individual variation and move beyond measures of screen time (e.g., Jensen et al., 2019; Pitt et al., 2021; Valkenburg et al., 2021; Verbeij et al., 2022).

**The Current Study**

The preceding twenty years of research has generated valuable insights into the changing nature and developmental impacts of adolescents' social media use. These insights are well-captured in recent review articles that summarize existing studies and take stock of overarching trends (e.g., Choukas-Bradley et al., 2023; Haidt & Twenge, ongoing; Hamilton, Nesi & Choukas-Bradley, 2022; Jensen, George, Russell & Odgers., 2019; Keles, McCrae & Grealish., 2020; Nesi, 2020; Odgers & Jensen, 2020; Odgers, Schueller & Ito, 2020; Orben, 2020; Valkenburg, Beyens, Meier, & Vanden Abeele, 2022). However, we lack an understanding of how researchers themselves view these trends. This understanding is needed to help set priorities for future research, especially research that seeks to transcend disciplinary siloes. The current study addresses this gap by asking the following questions:

> **RQ1:** How do researchers representing multiple disciplines describe trends in research on youth digital well-being?
>
> **RQ2:** What persistent gaps do these researchers identify as most pressing?
>
> **RQ3:** What areas of future research do researchers identify as priorities?

## Method

### Recruitment

To recruit a global, cross-disciplinary group of researchers, the first author began by drawing on her professional networks as a researcher who has studied youth digital well-being for nearly 20 years and whose research spans the fields of developmental science, education, and human-computer interaction (HCI). From there, we broadened our search using the "discipline" and "cited by" features in Google Scholar, exploring the citations within relevant articles, and searching within discipline-specific databases (e.g., ACM Digital Library for HCI, PsyNet for psychological science). To obtain a wider range of articles from around the world, we translated search terms using Google Translate and fed them back into Google Scholar. In addition, we used the interviews as an opportunity to ask researchers to identify others whom they thought should be part of this work. We identified and invited 46 scholars to participate in either an interview, the two-day convening, or both.

### Participants

Table 1 summarizes participants' fields, location, and whether they participated in an interview, the convening, or both. Participating researchers were primarily at the assistant and associate professor levels but included full professors and non-tenure-track research scientists. They came from universities and research institutes spanning 12 countries: Australia, Belgium, Canada, Colombia, Germany, Ireland, Italy, New Zealand, Spain, the Netherlands, the United Kingdom, and the United States. Within the United States, 11 states were represented: California, Florida, Massachusetts, New York, North Carolina, Oklahoma, Oregon, Pennsylvania, Rhode Island, Utah, and Washington.

Like the first author, several researchers identified with multiple fields. These included: developmental science, psychology, clinical psychology, clinical child psychology, adolescent psychology, educational psychology, education, the learning sciences, computer science, human-computer interaction (HCI), information science, science & technology studies (STS), sociology, anthropology, communication, global health, public health, and law.

This work was approved by the institutional review board of the first author's university. We have maintained the confidentiality of participants by referring to their participant ID. Where we do include names, we have secured the permission of the named researcher.

**Table 1 about here**

**Procedure**

During August-October 2021, the second author conducted interviews with 25 researchers via Zoom. The interviews lasted approximately 30-45 minutes and followed a semi-structured protocol. We asked participants to describe what youth digital well-being means to them; briefly summarize their research in the context of youth digital well-being; reflect on the current state of research in their field, including how research has addressed issues of diversity, equity, and inclusion; and identify areas that they feel should be prioritized in future research.

In April 2022, we hosted a two-day virtual convening with 26 researchers, 13 of whom had participated in the interviews (all interview participants were invited). The convening aimed to reach across disciplinary and geographical boundaries to learn how researchers in other disciplines and other countries are thinking about and researching youth digital well-being and to begin to form relationships with each other. We used the online platform gather.town to facilitate active conversations and informal interactions. Gather.town is a web-conferencing platform in which participants adopt avatars and move around a virtual space designed by the conference hosts.

The convening followed a World Café model (Brown, 2010). Small groups gathered at tables in the online convening space and were given approximately 25 minutes to discuss topics related to youth digital well-being. Topics included: surfacing field-specific research questions and methods; making research-practice connections; representation (who is doing the research, who is being researched); and identifying priorities for future research. After each discussion session, participants moved to a new table with a new group of people and a new topic. This format gave everyone the opportunity to meet and learn from a variety of people while connecting their ideas to widening circles of thought. We used the whiteboarding platform Miro to create a shared record of our conversations. These notes were supplemented by graduate student notetakers stationed at each table on Day 2 of the convening.

**Data Analysis**

We drew on Braun and Clarke's (2021) reflexive thematic analysis (TA) to analyze the interview and convening data. In contrast to more postpositivist approaches to coding and theme development (e.g., Boyatzis, 1998; Guest et al., 2011), reflexive TA is a more flexible, iterative approach (Braun & Clarke, 2019). Reflexive TA centers the researcher's subjectivity as an analytic resource and is therefore particularly fitting for our purposes. We are not merely writing about the researchers who study youth digital well-being; we count ourselves among these researchers. Indeed, three of the authors were also interview participants and all authors participated in the convening (all convening participants were invited to contribute to the current paper). To that end, we drew on our own expertise in youth digital well-being research to identify and synthesize key themes in our data and put them in conversation with published literature. We also drew on neo-ecological theory (Navarro & Tudge, 2022) by focusing our analysis on researchers' comments about the person- and context-specific factors associated with research on teens' social media experiences.

Our first stage of analysis involved an interplay between immersion in and distancing from the data (Braun & Clarke, 2022). To immerse ourselves in the data, three of the authors read through all the interview transcripts, interviewer notes, and convening documentation multiple times. We created distance from the data by recording, reflecting on, and discussing top-level impressions, cross-cutting patterns, and less common perspectives held by individual researchers. These discussions formed the basis of our coding as we developed and deepened our interpretation of the data. Consistent with reflexive TA, we did not create a codebook, instead treating emerging codes as provisional interpretations of the data (Braun & Clarke, 2022).

In the next stage of analysis, the remaining three authors joined the writing team and, together, we developed the initial codes into themes and discussed them in the context of published work; our own familiarity with and understanding of research on youth digital well-being; and our theoretical orientations grounded particularly in developmental science, clinical psychology, education, and human-computer interaction (Braun & Clarke, 2021). During our group conversations and memo writing, we grouped themes into three focal topics: *measurement*, *representation*, and *intervention*. Within each topic, we further organized the themes according to our three research questions: current state of research (RQ1), persistent gaps (RQ2), and future directions (RQ3). For instance, a theme within the *measurement* topic that addressed the current state of research (RQ1) was the perceived shift away from measuring the quantity of teens' social media use towards measuring the quality of teens' social media experiences. Table 2 summarizes all themes across focal topics and research questions.

Following best practices in thematic analysis (Nowell, Norris, White & Moules, 2017), we took measures to establish the trustworthiness of our findings (Elo et al., 2014; Lincoln & Guba, 1985). For instance, to establish credibility, we shared the findings with those participants who we quoted at length, asking them for feedback on the accuracy of our interpretations. In some cases, participants provided greater context for their statements, which we incorporated into the findings. To establish confirmability, we documented our analytic processes carefully, creating an audit trail with evidence of decisions made throughout our analysis (Nowell et al., 2017). We involved multiple researchers in the analysis, engaging in regular conversations in which we discussed our emerging interpretations of the data, co-constructed themes, and reflected together on each step of the research process.

**Study Limitations**

Although the multi-disciplinary, international focus is a strength of the work, we acknowledge that representation is far from exhaustive. In particular, there is a notable absence of researchers from Asian and African countries, and only one South American country is represented. This absence is partly attributable to time zone challenges and language barriers, which prevented some researchers in the Global South from participating. The absence is also reflective of where most researchers studying youth digital well-being are located. For instance, at the time of our convening, P30, whose work focuses on the Global South, was based in the UK. Thus, even though researchers based in this region are underrepresented in our sample, the perspectives captured in our findings include those from the Global South. In addition, although we achieved considerable disciplinary breadth, sub-disciplines within psychology as well as HCI are somewhat over-represented in our sample of researchers.

**Analysis and Discussion: Three Focal Topics in Youth Digital Well-Being Research**

This study is unique in its focus on exploring the perspectives of researchers. As such, participants' perspectives are connected to existing literature in a way that is not typical of most research focused on adolescents. Therefore, we present our findings and discussion as a single, synthesized section, placing participants' comments in direct conversation with existing literature relating to youth digital well-being.

We identified three focal topics discussed by interview and convening participants: measurement, representation, and intervention. Within each of these topics, participants discussed notable developments in their respective fields (RQ1), reflected on persistent gaps in the research base (RQ2), and offered ideas for future research priorities (RQ3). Table 2 summarizes the themes that we identified across the focal topics and research questions.

**Table 2 about here**

**Measurement**

**Developments related to measurement.** Participants identified a gradual shift in the way researchers measure the relationship between young people's technology use and well-being. Whereas early studies of youth's digital experiences relied heavily on measures of general use, such as amount, frequency, duration, or intensity (e.g., Ellison et al., 2007; Steinfield, Ellison &

Lampe, 2008), researchers are increasingly recognizing the importance of attending to the quality and nuance of these digital experiences.

P3 reflected on this shift away from measuring time spent on social media in the context of her own research. P3's research team tracked young people's use of digital media over the course of eight years, and their longitudinal data have brought important insights to our understanding of youth digital well-being (see Coyne et al., 2020). However, P3 acknowledged the shortcomings of the study's measures given it was designed more than a decade ago: *"It is hard and they're really basic measures. And so, by far, the limitation of our studies is that it's mostly time-based use, whereas [technology experiences are] so nuanced."*

P20 observed: *"I would love it if we could kind of lay aside the quantity question and move more to process and quality... Because I do think that there are a lot of interesting questions we can ask, in terms of what [teens are] doing online or via social media, and in some ways, our focus on screen time has not served us."* This quote reflects a sentiment shared by many participants—that it is time to move beyond time-based measures and focus instead on the quality of adolescents' social media experiences. As P14 summarized: *"it's… not just the amount of time that [youth] spend on technology, it's about quality of time and content and context."*

During the convening discussions, researchers shared approaches that their research teams were taking to measure the quality, content, and context of teens' social media experiences. These strategies shared a common characteristic in their shift from focusing on aggregate-level trends to person-specific dynamics. Person-specific analyses focus on individual trajectories or fluctuations that can help explain how changes in social media use affect changes in well-being, relative to an individual's baseline and across different scales of time. These analyses often build on theories regarding differential susceptibility to media effects (Valkenburg

& Peter, 2013), as well as broader work from the fields of developmental science and psychology on differential susceptibility, highlighting individual differences in susceptibility to the environment (e.g., see Belsky & Pluess, 2009; Ellis et al., 2011). Also relevant to these methodological approaches is Bronfenbrenner's Process-Person-Context-Time (PPCT) model that Navarro and Tudge (2022) have incorporated into neo-ecological theory.

Participants representing the fields of communications, developmental science, and psychology discussed two primary forms of person-specific analyses that they use in their research: (1) longitudinal analyses that examine within-person associations over long periods of time (e.g., Coyne et al., 2020; Gentile et al., 2011; Maheux et al., 2021), and (2) ecological momentary assessments that examine short-term associations between fluctuations in individuals' social media use and well-being (e.g., Pitt et al., 2021; Valkenburg et al., 2021; Verbeij et al., 2022). For example, we heard from Dr. Ine Beyens (P6), a research member of Project AWeSome (Adolescents, Well-being, Social Media) based at the University of Amsterdam about that team's experience sampling studies, which involve a large number of assessments within a short period of time. These researchers have conducted experience sampling studies that yield tens of thousands of data points, revealing sizeable differences in person-specific effects of time spent on social media on a range of developmental outcomes, such as self-esteem, friendship closeness, and distraction (e.g., Pouwels et al., 2021; Siebers, Beyens, Powels, & Valkenburg., 2022; Valkenburg et al., 2021).

Longitudinal and within-person analyses have helped researchers to uncover greater complexity in adolescents' social media experiences. At the same time, some participants acknowledged that many of these studies still rely on screen time as a dominant measure. During the convening, participants agreed on the need to develop scales to measure individual differences in adolescents' specific, subjective experiences with social media, and some described their efforts in this area. For example, within psychology, Choukas-Bradley, Zimmer-Gembeck, and others have developed and validated scales to assess the constructs of appearance-related social media consciousness (ASMC; Choukas-Bradley et al., 2020; Maheux et al., 2022b) and social media appearance preoccupation (Zimmer-Gembeck, Hawes & Pariz., 2021). ASMC has been found to significantly predict U.S. adolescents' body image concerns, depressive symptoms, and disordered eating, above and beyond overall time spent on social media, both cross-sectionally (Choukas-Bradley et al., 2020; Maheux et al., 2022a), and, in the case of depressive symptoms, longitudinally (Maheux et al., 2022b). Social media appearance preoccupation has been found to explain variance in Australian adolescents' symptoms of depression, social anxiety, appearance anxiety, and disordered eating, above and beyond a broad measure of social media use (Zimmer-Gembeck, Hawes, et al., 2021), with associations with appearance anxiety also documented longitudinally (Zimmer-Gembeck, Rudolph, et al., 2021).

In another example, from developmental science and educational psychology, Yang and colleagues developed scales to measure different forms of social comparison in the context of social media platforms. They found that judgmental social comparison on social media, which is competitive in nature and focuses on determining who is superior or inferior, was related to concurrent rumination and indirectly predicted poorer identity development. In contrast, informational social comparison, the goal of which is to be informed rather than to judge, was associated with adaptive identity processes (Yang, 2022; Yang, Holden, & Carter, 2018; Yang, Holden, Carter, & Webb, 2018).

**Persistent gaps, tensions, and paths forward**. Although research participants agreed that it is time to move beyond cross-sectional studies that use screen time as a dominant measure, they also observed that such research designs are still abundant in their respective fields (see Tang et al., 2021). By gathering data at a single moment in time, correlational studies miss opportunities to examine the likely reciprocal relationships between adolescents' social media use and their mental health outcomes (Keles et al., 2020; Odgers & Jensen, 2020; Orben, 2020). By using time as a proxy for social media use, these studies fail to capture the various experiences teens are having across different social media platforms (Choukas-Bradley et al., 2022; Hamilton et al., 2022; Orben, 2020; Valkenburg et al., 2022). Consistent with neo-ecological theory (Navarro & Tudge, 2022), these experiences are shaped by individual differences across teens, including the activities and people they engage with and their motives of use (Yang et al., 2021); differences in the features, affordances, and cultures found on different platforms (Davis, 2023); and the broader socio-cultural contexts of teens' social media use (Choukas-Bradley et al., 2022).

When it comes to accounting for individual variation in digital experiences, there is ongoing debate about the best way to approach person-specific analyses. For instance, Vuorre and colleagues at the Oxford Internet Institute (2022) have challenged the "personalized media effects paradigm" used by Project AWeSome researchers (see previous section) from both a methodological and theoretical perspective. They claim that the empirical results are not appropriately estimated and reported, and they cannot inform inferences about individuals. They argue that such study approaches ultimately contribute to confusing messages about the relationship between adolescents' social media use and well-being. This methodological disagreement highlights the challenges associated with conducting research that captures individual variation. Regardless of how this methodological disagreement unfolds, P1 observed that we need more than longitudinal experience sampling studies if we want to truly understand the impact of adolescents' social media use on mental health and well-being. *"You cannot address one subject only longitudinally or only with experience sampling data, or only with qualitative [data]—you have to do all of it I feel, to get the overview."*

Another persistent challenge discussed during the two-day convening is the considerable variation in the way researchers define and measure social media use (see also Keles et al., 2020; Valkenburg et al., 2022). P12 observed: *"It's very hard…to compare across studies or to come to any kind of consensus when everyone's asking things in a different way."* The fact that the convening represented multiple disciplines only highlighted these inconsistencies, including how rarely they are confronted due to disciplinary silos. Part of the challenge stems from the complexity inherent in adolescents' digital experiences. Platforms have different features, affordances, and cultures, which shape the quality of youth's experiences on them (Davis, 2023). As P15 noted, attending to this variation is important, but adds to the challenge of measurement: *"The nuances of social media really need to be focused on a little bit more, but it's not an easy thing to measure and to study."*

Participants offered several suggestions for directions forward when it comes to measurement and overall study design. First, although they acknowledged the presence of disciplinary silos, the conversations also highlighted the fact that many participants already collaborate with researchers in other fields. For instance, HCI researchers and clinical psychologists discussed collaborating on research exploring the design and effectiveness of technology-based mental health interventions (e.g., Dauden Roquet et al., 2022; Davis et al., 2023; Slovak et al., 2023). Through these types of collaborations, we are learning to communicate across disciplinary boundaries. Participants expressed optimism that increased cross-disciplinary work would encourage researchers to strive toward greater consistency in definition and measurement. They were also excited by the prospect of expanding what is possible in research on youth digital well-being. Returning to the example of the HCI/clinical psychologist collaboration, combining expertise in technology design with expertise in adolescent development and mental health intervention techniques promises to expand what is possible in the realm of effective technology-based mental health interventions. P19 asked: *"What would it mean to start shifting the learning [that happens in traditional therapeutic contexts] into the moments when you're supposed to apply the skills?"* P19 is referring here to research that seeks to use digital technologies to provide in-the-moment mental health interventions, such as through a mobile app.

Additional suggestions for strengthening the quality and interpretability of youth digital well-being research included moving towards an open science model, reporting effect sizes, and increasing the number of validated measures that are used across studies. In two separate interviews, researchers representing clinical psychology spoke about the need to focus on the specific mechanisms that lead to positive or negative outcomes, not just the social media activities themselves. Several researchers suggested making use of technologies such as sensors and eye tracking tools to help identify and document potential mechanisms.

**Representation**

**Developments related to representation.** Representation was a focal topic of discussion during the two-day convening. As research has begun to shift away from aggregate, time-based measures of social media use to person- and activity-specific measurements (see previous section), researchers observed a concurrent shift in focus towards increased representation of diverse adolescents, particularly with respect to race, gender and sexual identity, and cultural background. P15 offered an example from her research group, which is investigating youth's social media use in Australia, China, India, and Iran. *"We're looking forward to seeing how [social media and its effects] might differ in those cultures."*

Participants generally agreed that researchers have made progress toward understanding the distinct experiences of youth with marginalized identities. For example, studies from psychology, medicine, and public health have found that many sexual and gender minority (SGM) adolescents use social media to address unique developmental and identity-related needs, such as connecting with other SGM people, engaging in identity exploration processes, and seeking information and support regarding SGM identities, relationships, and health needs, which may improve mental health outcomes and identity acceptance among some SGM youth (Bates et al., 2020; Berger et al., 2021; Craig et al., 2021; McInroy, McCloskey, et al., 2019; Selkie et al., 2020).

Participants also observed greater research attention to the *intersection* of young people's marginalized identities. P9 reflected: *"I think we have seen a lot of research and researchers shift to more nuanced attention to differential susceptibility, surfacing questions about what is hard for whom, when and why, and taking approaches that just inherently reflect intersectionality and differential susceptibility."* For example, transgender youth of color may use social media to explore connections between their racial/ethnic and gender identities (Singh, 2013). Furthermore, SGM youth living in rural areas without a strong queer community presence may be especially likely to rely on social media to learn about SGM identities, connect with others, and reduce social isolation (Escobar-Viera et al., 2022; McInroy, Craig, et al., 2019; Paceley et al., 2022; Wike et al., 2022). On the other hand, for some SGM youth, social media experiences can negatively impact well-being, such as when youth are exposed to transphobic or homophobic content, harassment, or victimization (Berger et al., 2021; Selkie et al., 2020), or when they are shadow banned from social media platforms.

As researchers seek to understand the experiences of diverse adolescents, many are involving young people directly in research about them (e.g., Radesky et al., 2023; UNICEF, 2024). For instance, in their joint interview, P8 and P9 described a recent study in which they worked closely with a youth advisory group to interpret survey data they collected from teens across the United States (Weinstein & James, 2022). Through working with these teens, who represented a diversity of intersectional identities, the researchers were able to check their interpretations, identifying themes they had misinterpreted or even missed altogether. In our conversation with the researchers, P9 reflected: *"…young people's voices really need to be a feature of every part of the research process as much as possible… we don't just think about collecting data effectively from teens and young people, but we actually think about what might it look like if they really were partners in shaping the studies."*

Participants also observed an increase in the use of participatory and user-centered design methods when designing new technologies with and for adolescents. HCI has an established record of involving youth in participatory design research (e.g., Blandford et al., 2018; Druin, 1999; Scaife et al., 1997), but these methods have recently spread to other fields (Graham et al., 2019, Lyon, Brewer & Arean, 2020). For instance, an interdisciplinary group of social and computer scientists at Wellesley College, led by Dr. Linda Charmaraman (P14), created summer workshops to engage middle school students in the participatory design of more positive, inclusive online spaces. Charmaraman said of this work: *"[Youth] can be co-designers, co-creators. We can learn from them, they can teach us. We can teach them also that they have agency to have an impact in their own peer online spaces today."*

Such work underscores the value of including young people's voices in research on youth digital well-being (cf., also Psihogios, Lane-Fal, & Graham, 2022). In particular, it demonstrates how centering youth's perspectives in the study and design of new technologies can generate insights that adult researchers working alone would likely miss. These insights have bearing both on our understanding of youth's interactions with existing technologies and on the design of new technologies they will use in the future.

**Persistent gaps, tensions, and paths forward.** The relative paucity of cross-cultural work emerged as a major topic both in the group convening and individual interviews. Researchers pointed to the scarcity of research addressing the digital experiences of youth living in the Global South. Exceptions include P30's work exploring social media and adolescent well-being in the Global South (see also Andrade et al., 2022; Magis-Weinberg, Gys, Berger, Domoff, & Dahl, 2021; Sarwatay & Raman, 2022). During the convening, P30 observed that although the largest proportion of youth live in the Global South, their digital media practices are the least researched and understood (Ghai et al., 2022). She emphasized the importance of taking the time to understand the different cultural contexts that exist in the Global South countries, including regional, contextualized problems. The work of the Global Kids Online initiative, an international research project studying children's internet use globally, has been an important step in this direction. The initiative has generated reports on youth's online experiences in Argentina (Ravalli & Paolini, 2016), Chile (Cabello et al., 2019), Philippines (Tan, Estacio & Ylade, 2016), and South Africa (Phyfer, Burton, & Leoschut, 2016), among many others.

Even though research involving youth living in the Global North has begun to address diversity in meaningful ways (e.g., Cho et al., 2018; Delcourt et al., 2022; Watkins et al., 2018), participants observed there is still considerable work to be done. P14 reflected: *"We need more* *work on cultural differences, for instance racial ethnic identity or if English is the first language in your home, if you are an immigrant or not, if you have a household with different educational backgrounds or political backgrounds."* Other groups of youth underrepresented in youth digital well-being research include those from lower-income backgrounds and those with disabilities (Charmaraman et al., 2022). P14 went on to consider the care with which researchers should approach this type of research and how it may or may not exacerbate existing stereotypes: *"...for instance, if we know that Black and Latinx youth use online spaces and social media more than their white peers, a deficits lens would assume that this is a negative, alarming finding that youth of color are more vulnerable than their white counterparts, whereas an empowerment lens would want to investigate why and how youth of color are using online spaces more often in the first place."*

In discussing directions for future research, participants stressed the need to think beyond the diversity (or lack thereof) of youth who participate in their studies—researchers must also consider their own positionality. This includes how their identities and experiences shape the research questions they ask and the power dynamics that emerge between researcher and participants, especially if the latter belong to an underrepresented or marginalized group. We observed in the convening that most researchers who participated, as well as their collaborators, are white, cisgender women (including the majority of the authors of this paper). One researcher suggested that research teams make use of youth advisory groups when doing research across countries and cultures to help interpret and inform the direction of the research. As discussed in the previous section, involving youth directly in research has begun, but the researchers we spoke with agreed there is room for further development.

**Intervention**

      **Developments related to intervention.** Participants observed a greater amount of intervention-focused work emerging in recent years alongside an increased understanding of the opportunities and challenges of adolescents' digital experiences. Many of the researchers who participated in the interviews and two-day convening had direct experience with intervention-focused research. This work fell into three broad categories: (1) school-based curricula focused on supporting adolescents' positive digital experiences; (2) technology-based interventions that provide in-the-moment support for digital well-being; and (3) well-being interventions that use technology as a delivery mechanism for providing general mental health support.

      School-based interventions draw on long-standing traditions of media literacy education (see Jeong et al., 2012) that focus on developing the knowledge and skills needed to engage critically with mass media messages (e.g., the ability to recognize and challenge unrealistic body ideals in advertising), as well as digital literacy education focused on developing students' competence in both media use and production (Eyal & Te'eni-Harari, 2024). Digital well-being interventions typically seek to apply media and digital literacy pedagogies, as well as other techniques (e.g., cognitive behavioral therapy), to the challenges of digital media and technology (e.g., Paxton et al., 2022). For instance, P8 and P9 discussed how they partnered with Common Sense Media (CSM) and used their research to inform revisions to CSM's digital citizenship curriculum. As one example of a research-informed lesson, P8 described the "digital habits check-up," which *"invites youth to take stock of their digital habits and practices and reflect on what's serving them and what's detracting and what's challenging, and then choose something to work on."*

Digital well-being activities in the classroom vary in their focus, encompassing one or multiple aspects of technology use related to well-being. Example foci include cyberbullying, body image and diet culture, information quality and literacy, privacy and safety, and sexually explicit content (James & Weinstein, 2021; James, Weinstein & Mendoza, 2019). Although there is considerable variation in topic and pedagogical approach, a recent review of digital citizenship materials identified three common techniques across interventions: critical awareness, self-reflection, and strategies for behavior change (James et al., 2019; James & Weinstein, 2021). There is some emerging evidence that these techniques can support digital well-being (e.g., Bell et al., 2022), though more robust and large-scale evaluations of interventions in this domain are needed.

These interventions, many of which are initiated in a research context, are beginning to have real-world impact. In many countries, components of digital literacy have been incorporated into compulsory school curricula. For example, within the UK, recent revisions to the National Curriculum mean that all young people will be taught some knowledge and skills that are conducive to digital well-being, e.g., critiquing unrealistic appearance standards conveyed by social media (Department for Education, 2020). Common Sense Media's curriculum aligns with national and Common Core standards in the United States (Common Sense Media, 2019).

A second type of intervention, discussed particularly by HCI researchers, leverages technology design to support youth digital well-being. There is growing interest among HCI researchers in promoting personal well-being in people's daily technology use. Work in the areas of Positive Computing (Calvo & Peters, 2014), Positive Design (Desmet & Pohlmeyer, 2013), Positive Technology (Riva et al., 2012, 2020), and Digital Self-Control (Lukoff et al., 2022;

Roffarello & De Russis, 2021) illustrate this trend and the corresponding move away from user engagement as the dominant paradigm in interaction design (Mekler & Hornbaek, 2019).

Although most of this work focuses on general adult populations (Roffarello & De Russis, 2022), researchers are beginning to explore designs specific to adolescent well-being (e.g., Dauden Roquet et al., 2022; Davis et al., 2023; Galla et al., 2021). Examples from our convening and interviews include designing to support adolescent online safety and intentional social media use. P23, who focuses on designing for adolescent online safety, characterized her work as representative of a broader shift in the HCI field away from a deficit view of online participation and towards a more strengths-based approach: *"I've been trying to take a more resilience-based approach to online safety, of figuring out ways we can help teens self-regulate their behaviors and empower youth to protect themselves online."*

The third category of intervention discussed by participants uses technology as a vehicle for delivering general mental health interventions (Grist et al., 2017; Hamilton, Siegel & Carpenter, 2022; Schleider et al., 2020). Unlike the previous two categories of intervention, which treat technology as the primary target of intervention, this third approach leverages technology as an intervention delivery mechanism. For instance, in his research, P18 asks: *"What are the affordances that digital technology provides for solving some of the long-standing challenges we have faced with providing people with good options for both self-care of their mental health and well-being, as well as the delivery of mental health services?"* Allen and his colleagues are examining the use of mobile sensing as a method of tracking mental health status and mental health needs (e.g., Byrne et al., 2021; McNeilly et al., 2023). Such information could be used to detect when a teen may be at risk for attempting suicide. In another project, Allen and colleagues are developing an intervention that *"allows clinicians to access mobile sensing data* *to understand how their patients are doing between appointments…[which] allows the clinicians to build out therapy plans that then get pushed to the person via their phone."*

In another example from our interviews, Single Session Interventions (SSIs) are targeted, structured programs designed to have significant benefits to youth well-being following one engagement (Schleider et al., 2020). While SSIs can be delivered offline, web-based SSIs are designed for self-administration by youth, and so have wide potential reach and high scalability if made freely available (Schleider et al., 2020). P5, who explores SSIs in her work, said of the approach: *"What is more accessible than going directly to young people in the place they live already, and in the place they think and interact already? It seems so logical to create interventions for the environment that exists for young people rather than trying to make young people fit into [treatment] environments that have existed for decades…and didn't work very well."*

**Persistent gaps, tensions, and paths forward.** Participants across disciplines observed that more work is needed to develop interventions geared specifically to adolescents, particularly when it comes to technology-based interventions such as digital self-control tools and more general mental health resources (Hamilton et al., 2022; Roffarello & De Russis, 2022). Participants also noted the wide variation in intervention quality; few commercial interventions are rigorously evaluated before being made widely available to the public. Resources such as One Mind Psyber Guide (Schueller, 2022), which compiles and presents digital health resources reviewed by experts, including a Teen App Guide, are a step in the right direction.

Dr. Stephen Schueller (P7), who leads the One Mind Psyber Guide initiative, commented that there is a particular need for commercially available interventions that target diverse populations of adolescents: *"When you look at products that have reached maturity, we don't* *have a lot of mature products that are designed for diverse folks and audiences."* As an example, Schueller noted the paucity of Spanish-language products. He added: *"I think we have a ways to go to really build up a better evidence base about what works for people from diverse backgrounds and diverse populations, but I think the work's starting, which is great."*

P21 urged researchers to think beyond a Western therapy model when developing and targeting interventions for people from different cultures. This requires doing the work to understand different cultural contexts and what kinds of digital interventions fit best: *"I think that the field absolutely needs to be explicitly developing tools by and with people from diverse backgrounds and testing them with people from diverse backgrounds and seeking to understand in what ways they are helpful and useful."*

With respect to interventions developed in a research setting, disseminating them broadly to the youth who need them most remains a persistent challenge. P13 observed that research-based interventions need to be able to integrate into existing systems: *"I think to have a real impact on mental health, you need to design technologies that can integrate with either health or education systems, and you need to develop things which have the potential to operate at scale."* P21 emphasized the need to study the real-world impact of interventions, which is a core focus of implementation science (Eccles & Mittman, 2006). However, P21, along with several other participants, acknowledged that a major challenge with implementation is the mismatch between the slow, methodical pace of research and the "fail fast" ethos found in the tech industry. As a result, a promising intervention developed in a research context, with positive evidence from rigorous RCTs, might nevertheless fail by the time it reaches a broader audience, either because the technology has become outdated or its use in a real-world setting introduces variables that diminish its adoption and effectiveness (cf., for example Lyon & Koerner, 2016). Thus, future intervention work must seek ways of balancing academic rigor with practical usefulness.

## Implications and Conclusion

This work contributes a broad, multi-disciplinary overview of key insights, persistent gaps, and future paths in youth digital well-being research from the perspectives of researchers who are at the forefront of conducting this work. We engaged 38 researchers spanning 12 countries and multiple disciplines in one-on-one interviews and a two-day virtual convening that focused on mental health outcomes associated with adolescents' social media use. Informed by the neo-ecological theory (Navarro & Tudge, 2022), our study design and analysis focused on understanding the person- and context-specific factors associated with research on teens' social media experiences. We identified three focal topics related to the current state of this research landscape: measurement, representation, and intervention (see Table 2). Across these topics, participants observed that research has become increasingly fine-grained, with greater attention being paid to understanding the variety and quality of adolescents' social media experiences, as well as the individual characteristics and contexts of young people themselves. With this nuanced understanding, researchers are increasingly turning their attention to developing evidence-based interventions that support youth digital well-being.

At the same time, several gaps and tensions persist. For instance, despite the increased representation of diverse adolescents, there remains a scarcity of cross-cultural research and research focused on youth living in the Global South. With respect to intervention work, a persistent challenge is the difficulty of bringing evidence-based interventions to a broader audience due in part to the mismatch between the slow pace of academic research and the fast pace of the tech industry.

The paths forward summarized in Table 2 can be used to help funding agencies set priorities for supporting future research, and the insights generated from such research can be used to inform intervention and policy efforts aimed at supporting youth digital well-being. The current study points especially to the need for more cross-disciplinary, cross-cultural research. Cross-*disciplinary* research can help increase the clarity, consistency, and validity of definitions and measures. It can also contribute to more generative research and interventions, as researchers with different methodological skillsets and theoretical frameworks combine their knowledge to work on complex projects. Cross-*cultural* research can similarly contribute to the generativity of research on youth digital well-being by introducing a greater variety of perspectives, experiences, values, and commitments. In addition, research that spans cultures and geographies can help foreground researcher positionality and increase the diversity of both researchers and youth participants.

# References

Andrade, A. L. M., Scatena, A., De Oliveira Pinheiro, B., De Oliveira, W. A., Lopes, F. M., & De Micheli, D. (2022). Psychometric Properties of the Smartphone Addiction Inventory (SPAI-BR) in Brazilian Adolescents. *International Journal of Mental Health and Addiction*, *20*(5), 2690–2705. https://doi.org/10.1007/s11469-021-00542-x

Bates, A., Hobman, T., & Bell, B. T. (2020). "Let Me Do What I Please With It . . . Don't Decide My Identity For Me": LGBTQ+ Youth Experiences of Social Media in Narrative Identity Development. *Journal of Adolescent Research*, *35*(1), 51–83. https://doi.org/10.1177/0743558419884700

Baughan, A., Zhang, M. R., Rao, R., Lukoff, K., Schaadhardt, A., Butler, L. D., & Hiniker, A. (2022). "I Don't Even Remember What I Read": How Design Influences Dissociation on Social Media. *CHI Conference on Human Factors in Computing Systems*, 1–13. https://doi.org/10.1145/3491102.3501899

Bell, B. T., Taylor, C., Paddock, D., & Bates, A. (2022). Digital Bodies: A controlled evaluation of a brief classroom-based intervention for reducing negative body image among adolescents in the digital age. *British Journal of Educational Psychology*, *92*(1), e12449. https://doi.org/10.1111/bjep.12449

Belsky, J., & Pluess, M. (2009). Beyond diathesis stress: differential susceptibility to environmental influences. *Psychological bulletin*, *135*(6), 885.

Berger, M. N., Taba, M., Marino, J. L., Lim, M. S. C., Cooper, S. C., Lewis, L., Albury, K., Chung, K. S. K., Bateson, D., & Skinner, S. R. (2021). Social media's role in support networks among LGBTQ adolescents: A qualitative study. *Sexual Health*, *18*(5), 421–431. https://doi.org/10.1071/SH21110

Blandford, A., Gibbs, J., Newhouse, N., Perski, O., Singh, A., & Murray, E. (2018). Seven lessons for interdisciplinary research on interactive digital health interventions. *DIGITAL HEALTH*, *4*, 205520761877032. https://doi.org/10.1177/2055207618770325

Boyatzis, R. E. (1998). *Transforming qualitative information: Thematic analysis and code development*. Thousand Oaks, CA: Sage.

boyd, d. (2007). Why youth heart social network sites: The role of networked publics in teenage social life. In D. Buckingham (Ed.), *Youth, identity, and digital media* (pp. 119-142). Cambridge, MA: MIT Press.

boyd, d. (2014.). *It's complicated: The social lives of networked teens*. New Haven, CT: Yale University Press.

Braun, V., & Clarke, V. (2019). Reflecting on reflexive thematic analysis. *Qualitative Research in Sport, Exercise and Health*, *11*(4), 589–597. https://doi.org/10.1080/2159676X.2019.1628806

Braun, V., & Clarke, V. (2021). One size fits all? What counts as quality practice in (reflexive) thematic analysis? *Qualitative Research in Psychology*, *18*(3), 328–352.

Braun, V., & Clarke, V. (2022). Conceptual and design thinking for thematic analysis. *Qualitative Psychology*, *9*, 3–26. https://doi.org/10.1037/qup0000196

Brown, J. (2010). *The world café: Shaping our futures through conversations that matter*. ReadHowYouWant. com.

Byrne, M. L., Lind, M. N., Horn, S. R., Mills, K. L., Nelson, B. W., Barnes, M. L., ... & Allen, N. B. (2021). Using mobile sensing data to assess stress: Associations with perceived and lifetime stress, mental health, sleep, and inflammation. *Digital Health*, *7*, 20552076211037227.

Cabello, P., Claro, M., Lazcano, D. Cabello-Hutt, T., Antezana, L., and Ochoa, J.M. (2019). Global Kids Online Chile: Chilean children's internet use and online activities, a brief report. Available at: http://globalkidsonline.net/chile/

Calvo, R. A., & Peters, D. (2014). *Positive computing: Technology for wellbeing and human potential*. MIT Press.

Charmaraman, L., & Grevet Delcourt, C. (2021). Prototyping for Social Wellbeing with Early Social Media Users: Belonging, Experimentation, and Self-Care. *Proceedings of the 2021 CHI Conference on Human Factors in Computing Systems*, 1–15. https://doi.org/10.1145/3411764.3445332

Charmaraman, L., Hernandez, J. M., & Hodes, R. (2022). Marginalized and Understudied Populations Using Digital Media. In J. Nesi, E. H. Telzer, & M. J. Prinstein (Eds.), *Handbook of Adolescent Digital Media Use and Mental Health* (1st ed., pp. 188–214). Cambridge University Press. https://doi.org/10.1017/9781108976237.011

Cho, A. (2018). Default publicness: Queer youth of color, social media, and being outed by the machine. *New Media & Society*, *20*(9), 3183–3200. https://doi.org/10.1177/1461444817744784

Choukas-Bradley, S., Kilic, Z., Stout, C. D., & Roberts, S. R. (2023). Perfect Storms and Double-Edged Swords: Recent Advances in Research on Adolescent Social Media Use and Mental Health. *Advances in Psychiatry and Behavioral Health*, S2667382723000078. https://doi.org/10.1016/j.ypsc.2023.03.007

Choukas-Bradley, S., Nesi, J., Widman, L., & Galla, B. M. (2020). The Appearance-Related Social Media Consciousness Scale: Development and validation with adolescents. *Body Image*, *33*, 164–174. https://doi.org/10.1016/j.bodyim.2020.02.017

Choukas-Bradley, S., Roberts, S. R., Maheux, A. J., & Nesi, J. (2022). The Perfect Storm: A Developmental–Sociocultural Framework for the Role of Social Media in Adolescent Girls' Body Image Concerns and Mental Health. *Clinical Child and Family Psychology Review*. https://doi.org/10.1007/s10567-022-00404-5

Common Sense Media (2019). *New digital citizenship standards alignment*. Available at: https://www.commonsense.org/education/new-digital-citizenship-standards-alignment-2019

Costello, M., & Hawdon, J. (2020). Hate Speech in Online Spaces. In T. J. Holt & A. M. Bossler (Eds.), *The Palgrave Handbook of International Cybercrime and Cyberdeviance* (pp. 1397–1416). Springer International Publishing. https://doi.org/10.1007/978-3-319-78440-3_60

Coyne, S. M., Rogers, A. A., Zurcher, J. D., Stockdale, L., & Booth, M. (2020). Does time spent using social media impact mental health?: An eight year longitudinal study. *Computers in Human Behavior*, *104*, 106160. https://doi.org/10.1016/j.chb.2019.106160

Craig, S. L., Eaton, A. D., McInroy, L. B., Leung, V. W. Y., & Krishnan, S. (2021). Can Social Media Participation Enhance LGBTQ+ Youth Well-Being? Development of the Social Media Benefits Scale. *Social Media + Society*, *7*(1), 205630512198893. https://doi.org/10.1177/2056305121988931

Daudén Roquet, C., Theofanopoulou, N., Freeman, J. L., Schleider, J., Gross, J. J., Davis, K., Townsend, E., & Slovak, P. (2022). Exploring Situated & Embodied Support for Youth's Mental Health: Design Opportunities for Interactive Tangible Device. *CHI Conference on Human Factors in Computing Systems*, 1–16.

Davis, K. (2010). Coming of age online: The developmental underpinnings of girls' blogs. *Journal of Adolescent Research, 25* (1), 145-171.

Davis, K. (2012a). Friendship 2.0: Adolescents' experiences of belonging and self-disclosure online. *Journal of Adolescence, 35* (6), 1527-1536.

Davis, K. (2012b). Tensions of identity in a networked era: Young people's perspectives on the risks and rewards of online self-expression. *New Media & Society, 14* (4), 634-651.

Davis, K. (2014). Youth identities in a digital age: The anchoring role of friends in youth's approaches to online identity expression. In A. Bennett and B. Robards (Eds.), *Mediated youth cultures: The internet, belonging, and new cultural configurations* (pp.11-25). New York: Palgrave Macmillan.

Davis, K. (2023). *Technology's child: Digital media's role in the ages and stages of growing up*. Cambridge, MA: The MIT Press.

Davis, K., Slovak, P., Landesman, R., Pitt, C., Ghajar, A., Schleider, J.L., Kawas, S., Perez Portillo, A..G., & Kuhn, N.S. (2023). Supporting teens' intentional social media use through interaction design: An exploratory proof-of-concept study. *Proceedings of the ACM SIGCHI Conference on Interaction Design and Children (IDC '23)*. New York: ACM Press.

Delcourt, C. G., Charmaraman, L., Durrani, S., Gu, Q., & Xiao, L. F. (2022, April). Innovating Novel Online Social Spaces with Diverse Middle School Girls: Ideation and Collaboration in a Synchronous Virtual Design Workshop. In *Proceedings of the 2022 CHI Conference on Human Factors in Computing Systems* (pp. 1-16).

Desmet, P. M., & Pohlmeyer, A. E. (2013). Positive design: An introduction to design for subjective well-being. *International Journal of Design*, *7*(3).

Druin, A. (1999). Cooperative inquiry: Developing new technologies for children with children. *Proceedings of the SIGCHI Conference on Human Factors in Computing Systems the CHI Is the Limit - CHI '99*, 592–599. https://doi.org/10.1145/302979.303166

Eccles, M. P., & Mittman, B. S. (2006). Welcome to Implementation Science. *Implementation Science*, *1*(1), 1, 1748-5908-1–1. https://doi.org/10.1186/1748-5908-1-1

Escobar-Viera, C. G., Choukas-Bradley, S., Sidani, J., Maheux, A. J., Roberts, S. R., & Rollman, B. L. (2022). Examining Social Media Experiences and Attitudes Toward Technology-Based Interventions for Reducing Social Isolation Among LGBTQ Youth Living in Rural United States: An Online Qualitative Study. *Frontiers in Digital Health*, *4*, 900695. https://doi.org/10.3389/fdgth.2022.900695

Ellis, B. J., Boyce, W. T., Belsky, J., Bakermans-Kranenburg, M. J., & Van IJzendoorn, M. H. (2011). Differential susceptibility to the environment: An evolutionary–neurodevelopmental theory. *Development and psychopathology*, *23*(1), 7-28.

Ellison, N. B., Steinfield, C., & Lampe, C. (2007). The Benefits of Facebook "Friends:" Social Capital and College Students' Use of Online Social Network Sites. *Journal of Computer-Mediated Communication*, *12*(4), 1143–1168. https://doi.org/10.1111/j.1083-6101.2007.00367.x

Elo, S., Kääriäinen, M., Kanste, O., Pölkki, T., Utriainen, K., & Kyngäs, H. (2014). Qualitative Content Analysis: A Focus on Trustworthiness. *SAGE Open*, *4*(1), 215824401452263. https://doi.org/10.1177/2158244014522633

Galla, B. M., Choukas-Bradley, S., Fiore, H. M., & Esposito, M. V. (2021). Values-Alignment Messaging Boosts Adolescents' Motivation to Control Social Media Use. *Child Development*, *92*(5), 1717–1734. https://doi.org/10.1111/cdev.13553

Gentile, D. A., Choo, H., Liau, A., Sim, T., Li, D., Fung, D., & Khoo, A. (2011). Pathological video game use among youths: A two-year longitudinal study. *Pediatrics*, *127*(2), e319-e329.

Ghai, S., Magis-Weinberg, L., Stoilova, M., Livingstone, S., & Orben, A. (2022). Social media and adolescent well-being in the Global South. *Current Opinion in Psychology*, *46*, 101318. https://doi.org/10.1016/j.copsyc.2022.101318

Giumetti, G. W., & Kowalski, R. M. (2022). Cyberbullying via social media and well-being. *Current Opinion in Psychology*, *45*, 101314. https://doi.org/10.1016/j.copsyc.2022.101314

Graham, A. K., Wildes, J. E., Reddy, M., Munson, S. A., Barr Taylor, C., & Mohr, D. C. (2019). User-centered design for technology-enabled services for eating disorders. *International Journal of Eating Disorders*, *52*(10), 1095–1107. https://doi.org/10.1002/eat.23130

Granic, I., Morita, H., & Scholten, H. (2020). Beyond Screen Time: Identity Development in the Digital Age. *Psychological Inquiry*, *31*(3), 195–223. https://doi.org/10.1080/1047840X.2020.1820214

Grist, R., Porter, J., & Stallard, P. (2017). Mental Health Mobile Apps for Preadolescents and Adolescents: A Systematic Review. *Journal of Medical Internet Research*, *19*(5), e176. https://doi.org/10.2196/jmir.7332

Guest, G., MacQueen ,K. and E. Namey, E. (2011). *Applied thematic analysis.* Thousand Oaks, CA: Sage.

Haidt, J., & Twenge, J. (ongoing). *Adolescent mood disorders since 2010: A collaborative review.* https://docs.google.com/document/d/1diMvsMeRphUH7E6D1d_J7R6WbDdgnzFHDHPx9HXzR5o/edit#

Hamilton, J. L., Nesi, J., & Choukas-Bradley, S. (2022). Re-examining adolescent social media use and socioemotional well-being through the lens of the COVID-19 pandemic: A theoretical review and directions for future research. *Perspectives on Psychological Science: A Journal of the Association for Psychological Science*, *17*(3), 662–679. https://doi.org/10.1177/17456916211014189

Hamilton, J. L., Siegel, D. M., & Carper, M. M. (2022). Digital Media Interventions for Adolescent Mental Health. In E. H. Telzer, J. Nesi, & M. J. Prinstein (Eds.), *Handbook of Adolescent Digital Media*

*Use and Mental Health* (pp. 389–416). Cambridge University Press. https://doi.org/10.1017/9781108976237.021

Ito, M., Baumer, S., Bittanti, M., boyd, d., Cody, R., Herr-Stephenson, B., et al. (2009). *Hanging out, messing around, and geeking out: Kids living and learning with new media.* Cambridge, MA: The MIT Press.

James, C., Weinstein, Emily, & Mendoza, Kelly. (2019). *Teaching Digital Citizens in Today's World: Research and Insights Behind the Common Sense K–12 Digital Citizenship Curriculum*. Common Sense Media.

James, C., & Weinstein, E. (2021). Children as Digital Citizens: Insights from classroom research with digital dilemmas. *Journal of E-Learning and Knowledge Society*, 5-7 Pages. https://doi.org/10.20368/1971-8829/1135583

Jensen, M., George, M., Russell, M., & Odgers, C. (2019). Young Adolescents' Digital Technology Use and Mental Health Symptoms: Little Evidence of Longitudinal or Daily Linkages. *Clinical Psychological Science : A Journal of the Association for Psychological Science*, *7*(6), 1416–1433. https://doi.org/10.1177/2167702619859336

Jeong, S. H., Cho, H., & Hwang, Y. (2012). Media literacy interventions: A meta-analytic review. *Journal of Communication*, *62*(3), 454-472.

Karim, S., Choukas-Bradley, S., Radovic, A., Roberts, S. R., Maheux, A. J., & Escobar-Viera, C. G. (2022). Support over Social Media among Socially Isolated Sexual and Gender Minority Youth in Rural U.S. during the COVID-19 Pandemic: Opportunities for Intervention Research. *International Journal of Environmental Research and Public Health*, *19*(23), 15611. https://doi.org/10.3390/ijerph192315611

Keles, B., McCrae, N., & Grealish, A. (2020). A systematic review: The influence of social media on depression, anxiety and psychological distress in adolescents. *International Journal of Adolescence and Youth*, *25*(1), 79–93.

Kleemans, M., Daalmans, S., Carbaat, I., & Anschütz, D. (2018). Picture perfect: The direct effect of manipulated Instagram photos on body image in adolescent girls. *Media Psychology*, *21*(1), 93–110.

Lenhart, A., Madden, M., Rankin Macgill, A., & Smith, A. (2007, December 19). Teens and social media: The use of social media gains a greater foothold in teen life as they embrace the conversational nature of interactive online media (PEW Internet & American Life Project). Retrieved December 20, 2023, from https://www.pewinternet.org/wp-content/uploads/sites/9/media/Files/Reports/2007/PIP_Teens_Social_Media_Final.pdf.pdf

Lincoln S. Y., Guba E. G. (1985). *Naturalistic inquiry*. Thousand Oaks, CA: Sage.

Lopes, L. S., Valentini, J. P., Monteiro, T. H., Costacurta, M. C. de F., Soares, L. O. N., Telfar-Barnard, L., & Nunes, P. V. (2022). Problematic Social Media Use and Its Relationship with Depression or Anxiety: A Systematic Review. *Cyberpsychology, Behavior, and Social Networking*, *25*(11), 691–702. https://doi.org/10.1089/cyber.2021.0300

Lyon, A. R., Brewer, S. K., & Areán, P. A. (2020). Leveraging human-centered design to implement modern psychological science: Return on an early investment. *American Psychologist*, *75*(8), 1067–1079. https://doi.org/10.1037/amp0000652

Lyon, A. R., & Koerner, K. (2016). User-centered design for psychosocial intervention development and implementation. *Clinical Psychology: Science and Practice*, *23*(2), 180–200. https://doi.org/10.1111/cpsp.12154

Magis-Weinberg, L., Ballonoff Suleiman, A., & Dahl, R. E. (2021a). Context, Development, and Digital Media: Implications for Very Young Adolescents in LMICs. *Frontiers in Psychology*, *12*, 632713. https://doi.org/10.3389/fpsyg.2021.632713

Magis-Weinberg, L., Gys, C. L., Berger, E. L., Domoff, S. E., & Dahl, R. E. (2021b). Positive and Negative Online Experiences and Loneliness in Peruvian Adolescents During the COVID-19 Lockdown. *Journal of Research on Adolescence*, *31*(3), 717–733. https://doi.org/10.1111/jora.12666

Maheux, A. J., Nesi, J., Galla, B. M., Roberts, S. R., & Choukas-Bradley, S. (2021). #Grateful: Longitudinal Associations Between Adolescents' Social Media Use and Gratitude During the COVID-19 Pandemic. *Journal of Research on Adolescence*, *31*(3), 734–747. https://doi.org/10.1111/jora.12650

Maheux, A. J., Roberts, S. R., Nesi, J., Widman, L., & Choukas-Bradley, S. (2022a). Longitudinal associations between appearance-related social media consciousness and adolescents' depressive symptoms. *Journal of Adolescence*, *94*(2), 264–269. https://doi.org/10.1002/jad.12009

Maheux, A. J., Roberts, S. R., Nesi, J., Widman, L., & Choukas-Bradley, S. (2022b). Psychometric properties and factor structure of the appearance-related social media consciousness scale among emerging adults. *Body Image*, *43*, 63–74. https://doi.org/10.1016/j.bodyim.2022.08.002

McCabe, M. A., Prinstein, M. J., Alvord, M. K., Bounds, D. T., Charmaraman, L., Choukas-Bradley, S., Espelage, D. L., Goodman, J. A., Hamilton, J. L., Nesi, J., Tynes, B. M., Ward, L. M., & Magis-Weinberg, L. (2023). *Health advisory on social media use in adolescence*. American Psychological Association. https://www.apa.org/topics/social-media-internet/health-advisory-adolescent-social-media-use

McInroy, L. B., Craig, S. L., & Leung, V. W. Y. (2019). Platforms and Patterns for Practice: LGBTQ+ Youths' Use of Information and Communication Technologies. *Child and Adolescent Social Work Journal*, *36*(5), 507–520. https://doi.org/10.1007/s10560-018-0577-x

McInroy, L. B., McCloskey, R. J., Craig, S. L., & Eaton, A. D. (2019). LGBTQ+ Youths' Community Engagement and Resource Seeking Online versus Offline. *Journal of Technology in Human Services*, *37*(4), 315–333. https://doi.org/10.1080/15228835.2019.1617823

McNeilly, E. A., Mills, K. L., Kahn, L. E., Crowley, R., Pfeifer, J. H., & Allen, N. B. (2023). Adolescent Social Communication Through Smartphones: Linguistic Features of Internalizing Symptoms and Daily Mood. *Clinical Psychological Science*, 216770262211251. https://doi.org/10.1177/21677026221125180

Mekler, E. D., & Hornbæk, K. (2019). A framework for the experience of meaning in human-computer interaction. *Proceedings of the 2019 CHI Conference on Human Factors in Computing Systems*, 1–15.

Meier, A., & Reinecke, L. (2021). Computer-Mediated Communication, Social Media, and Mental Health: A Conceptual and Empirical Meta-Review. *Communication Research*, *48*(8), 1182–1209. https://doi.org/10.1177/0093650220958224

Middaugh, E., Clark, L. S., & Ballard, P. J. (2017). Digital Media, Participatory Politics, and Positive Youth Development. *Pediatrics*, *140*(Supplement_2), S127–S131. https://doi.org/10.1542/peds.2016-1758Q

Mkhize, S., Nunlall, R., & Gopal, N. (2020). An examination of social media as a platform for cyber-violence against the LGBT+ population. *Agenda*, *34*(1), 23–33. https://doi.org/10.1080/10130950.2019.1704485

Navarro, J. L., & Tudge, J. R. H. (2023). Technologizing Bronfenbrenner: Neo-ecological Theory. *Current Psychology*, *42*(22), 19338–19354. https://doi.org/10.1007/s12144-022-02738-3

Nesi, J. (2020). The Impact of Social Media on Youth Mental Health: Challenges and Opportunities. *North Carolina Medical Journal*, *81*(2), 116–121. https://doi.org/10.18043/ncm.81.2.116

Nesi, J., & Prinstein, M. J. (2015). Using Social Media for Social Comparison and Feedback-Seeking: Gender and Popularity Moderate Associations with Depressive Symptoms. *Journal of Abnormal Child Psychology*, *43*(8), 1427–1438. https://doi.org/10.1007/s10802-015-0020-0

Nesi, J., Choukas-Bradley, S., & Prinstein, M. J. (2018). Transformation of adolescent peer relations in the social media context: Part 1—a theoretical framework and application to dyadic peer relationships. *Clinical child and family psychology review*, *21*(3), 267-294.

Nick, E. A., Kilic, Z., Nesi, J., Telzer, E. H., Lindquist, K. A., & Prinstein, M. J. (2022). Adolescent Digital Stress: Frequencies, Correlates, and Longitudinal Association With Depressive Symptoms. *Journal of Adolescent Health*, *70*(2), 336–339. https://doi.org/10.1016/j.jadohealth.2021.08.025

Nowell, L. S., Norris, J. M., White, D. E., & Moules, N. J. (2017). Thematic analysis: Striving to meet the trustworthiness criteria. *International journal of qualitative methods*, *16*(1), 1609406917733847.

Odgers, C. L., & Jensen, M. (2020). Adolescent Mental Health in the Digital Age: Facts, Fears and Future Directions. *Journal of Child Psychology and Psychiatry, and Allied Disciplines*, *61*(3), 336–348. https://doi.org/10.1111/jcpp.13190

Odgers, C. L., Schueller, S. M., & Ito, M. (2020). Screen Time, Social Media Use, and Adolescent Development. *Annual Review of Developmental Psychology*, *2*(1), 485–502. https://doi.org/10.1146/annurev-devpsych-121318-084815

Office of the Surgeon General (OSG), 2023. Social Media and Youth Mental Health.

Orben, A. (2020a). Teenagers, screens and social media: A narrative review of reviews and key studies. *Social Psychiatry and Psychiatric Epidemiology*, *55*(4), 407–414. https://doi.org/10.1007/s00127-019-01825-4

Paceley, M. S., Goffnett, J., Sanders, L., & Gadd-Nelson, J. (2022). "Sometimes You Get Married on Facebook": The Use of Social Media Among Nonmetropolitan Sexual and Gender Minority Youth. *Journal of Homosexuality*, *69*(1), 41–60. https://doi.org/10.1080/00918369.2020.1813508

Paxton, S. J., McLean, S. A., & Rodgers, R. F. (2022). "My critical filter buffers your app filter": Social media literacy as a protective factor for body image. *Body image*, *40*, 158-164.

Pew Research Center (December 2023), "Teens, Social Media and Technology 2023." Available at: https://www.pewresearch.org/internet/2023/12/11/teens-social-media-and-technology-2023/

Phyfer, J., Burton, P. & Leoschut, L. (2016). *Global Kids Online South Africa: Barriers, opportunities and risks. A glimpse into South African children's internet use and online activities. Technical Report.* Cape Town: Centre for Justice and Crime Prevention. Available from: www.globalkidsonline.net/south-africa

Pitt, C., Hock, A., Zelnick, L., & Davis, K. (2021). The Kids Are / Not / Sort of All Right*. *Proceedings of the 2021 CHI Conference on Human Factors in Computing Systems*, 1–14. https://doi.org/10.1145/3411764.3445541

Pouwels, J. L., Valkenburg, P. M., Beyens, I., van Driel, I. I., & Keijsers, L. (2021). Social media use and friendship closeness in adolescents' daily lives: An experience sampling study. *Developmental Psychology*, *57*(2), 309. https://doi.org/10.1037/dev0001148

Psihogios, A. M., Lane-Fall, M. B., & Graham, A. K. (2022). Adolescents Are Still Waiting on a Digital Health Revolution: Accelerating Research-to-Practice Translation Through Design for Implementation. *JAMA Pediatrics*, *176*(6), 545. https://doi.org/10.1001/jamapediatrics.2022.0500

Radesky, J., Weeks, H. M., Schaller, A., Robb, M., Mann, S., & Lenhart, A. (2023). *Constant Companion: A Week in the Life of a Young Person's Smartphone Use*. Common Sense Media. https://www.commonsensemedia.org/sites/default/files/research/report/2023-cs-smartphone-research-report_final-for-web.pdf

Ravalli, M. and Paoloni, P. (2016). *Global Kids Online Argentina: Research study on the perceptions and habits of children and adolescents on the use of technologies, the internet and social media.* Buenos Aires: UNICEF Argentina, www.globalkidsonline/argentina

Riva, G., Baños, R. M., Botella, C., Wiederhold, B. K., & Gaggioli, A. (2012). Positive technology: Using interactive technologies to promote positive functioning. *Cyberpsychology, Behavior, and Social Networking*, *15*(2), 69–77.

Riva, G., Mantovani, F., & Wiederhold, B. K. (2020). Positive Technology and COVID-19. *Cyberpsychology, Behavior, and Social Networking*.

Roffarello, A. M., & De Russis, L. (2021). Coping with Digital Wellbeing in a Multi-Device World. *Proceedings of the 2021 CHI Conference on Human Factors in Computing Systems*, 1–14. https://doi.org/10.1145/3411764.3445076

Roffarello, A. M., & De Russis, L. (2022). Achieving Digital Wellbeing Through Digital Self-Control Tools: A Systematic Review and Meta-Analysis. *ACM Transactions on Computer-Human Interaction*.

Sala, A., Porcaro, L., & Gómez, E. (2024). Social Media Use and adolescents' mental health and well-being: An umbrella review. *Computers in Human Behavior Reports*, *14*, 100404. https://doi.org/10.1016/j.chbr.2024.100404

Sarwatay, D., & Raman, U. (2022). Everyday negotiations in managing presence: Young people and social media in India. *Information, Communication & Society*, *25*(4), 536–551. https://doi.org/10.1080/1369118X.2021.1988129

Scaife, M., Rogers, Y., Aldrich, F., & Davies, M. (1997). Designing for or designing with? Informant design for interactive learning environments. *Proceedings of the ACM SIGCHI Conference on Human Factors in Computing Systems*, 343–350. https://doi.org/10.1145/258549.258789

Schleider, J. L., Dobias, M. L., Sung, J. Y., & Mullarkey, M. C. (2020). Future Directions in Single-Session Youth Mental Health Interventions. *Journal of Clinical Child & Adolescent Psychology*, *49*(2), 264–278. https://doi.org/10.1080/15374416.2019.1683852

Schouten, A. P., Valkenburg, P. M., & Peter, J. (2007). Precursors and underlying processes of adolescents' online self-disclosure: Developing and testing an 'internet-attribute-perception' model. *Media Psychology, 10*(2), 292-315.

Schueller, S. M. (2022). Navigating the Digital Mental Health Landscape: One Mind PsyberGuide. *Signal*, 08.

Selkie, E., Adkins, V., Masters, E., Bajpai, A., & Shumer, D. (2020). Transgender Adolescents' Uses of Social Media for Social Support. *Journal of Adolescent Health*, *66*(3), 275–280. https://doi.org/10.1016/j.jadohealth.2019.08.011

Siebers, T., Beyens, I., Pouwels, J. L., & Valkenburg, P. M. (2022). Social Media and Distraction: An Experience Sampling Study among Adolescents. *Media Psychology*, *25*(3), 343–366. https://doi.org/10.1080/15213269.2021.1959350

Singh, A. A. (2013). Transgender Youth of Color and Resilience: Negotiating Oppression and Finding Support. *Sex Roles*, *68*(11–12), 690–702. https://doi.org/10.1007/s11199-012-0149-z

Slovak, P., Antle, A. N., Theofanopoulou, N., Roquet, C. D., Gross, J. J., & Isbister, K. (2023). Designing for emotion regulation interventions: An agenda for HCI theory and research. *ACM Transactions on Computer-Human Interaction*, *30*(1), 1–51.

Steele, R. G., Hall, J. A., & Christofferson, J. L. (2020). Conceptualizing Digital Stress in Adolescents and Young Adults: Toward the Development of an Empirically Based Model. *Clinical Child and Family Psychology Review*, *23*(1), 15–26. https://doi.org/10.1007/s10567-019-00300-5

Steinfield, C., Ellison, N. B., & Lampe, C. (2008). Social capital, self-esteem, and use of online social network sites: A longitudinal analysis. *Journal of Applied Developmental Psychology*, *29*(6), 434–445. https://doi.org/10.1016/j.appdev.2008.07.002

Stiffler, L. (2023, October 24). Washington and 41 state AGs sue social media giant Meta for its alleged harm of youth. *GeekWire*. https://www.geekwire.com/2023/washington-and-41-states-sue-social-media-giant-meta-for-its-alleged-harm-of-youth/

Tan, M., Estacio, L., and Ylade, M. (2016). *Global Kids Online in the Philippines. Country Report.* Manila: University of the Philippines Manila. Available from: www.globalkidsonline/philippines

Tang, S., Werner-Seidler, A., Torok, M., Mackinnon, A. J., & Christensen, H. (2021). The relationship between screen time and mental health in young people: A systematic review of longitudinal studies. *Clinical Psychology Review*, *86*, 102021. https://doi.org/10.1016/j.cpr.2021.102021

Tanksley, T. C. (2019). *Race, education and# BlackLivesMatter: How social media activism shapes the educational experiences of Black college-age women* [Dissertation]. University of California, Los Angeles.

Tomczyk, S., & Hoferichter, F. (2022). Associations between social media use, psychological stress, well-being, and alpha-amylase levels in adolescents. *Journal of Stress, Trauma, Anxiety, and Resilience (J-STAR)*, *1*(2), 26–37.

Tynes, B. M., English, D., Del Toro, J., Smith, N. A., Lozada, F. T., & Williams, D. R. (2020). Trajectories of online racial discrimination and psychological functioning among African American and Latino adolescents. *Child Development*, *91*(5), 1577-1593.

UNICEF (2024, April). *Responsible Innovation in Technology for Children: Digital technology, play and child well-being*. Florence: UNICEF Innocenti – Global Office of Research and Foresight.

Valkenburg, P., Beyens, I., Pouwels, J. L., van Driel, I. I., & Keijsers, L. (2021). Social Media Use and Adolescents' Self-Esteem: Heading for a Person-Specific Media Effects Paradigm. *Journal of Communication*, *71*(1), 56–78. https://doi.org/10.1093/joc/jqaa039

Valkenburg, P. M., Beyens, I., Meier, A., & Vanden Abeele, M. M. P. (2022). Advancing our understanding of the associations between social media use and well-being. *Current Opinion in Psychology*, *47*, 101357. https://doi.org/10.1016/j.copsyc.2022.101357

Valkenburg, P. M., & Peter, J. (2013). The Differential Susceptibility to Media Effects Model: Differential Susceptibility to Media Effects Model. *Journal of Communication*, *63*(2), 221–243. https://doi.org/10.1111/jcom.12024

Verbeij, T., Pouwels, J. L., Beyens, I., & Valkenburg, P. M. (2022). Experience sampling self-reports of social media use have comparable predictive validity to digital trace measures. *Scientific Reports*, *12*(1), Article 1. https://doi.org/10.1038/s41598-022-11510-3

Vuorre, M., Johannes, N., & Przybylski, A. K. (2022). *Three objections to a novel paradigm in social media effects research*. PsyArXiv. https://doi.org/10.31234/osf.io/dpuya

Watkins, S. C., Cho, A., Lombana-Bermudez, A., Shaw, V., Vickery, J. R., & Weinzimmer, L. (2018). *The Digital Edge: How Black and Latino Youth Navigate Digital Inequality*. NYU Press.

Weinstein, E., & James, C. (2022). *Behind their screens: What teens are facing (and adults are missing)*. MIT Press.

Wike, T. L., Bouchard, L. M., Kemmerer, A., & Yabar, M. P. (2022). Victimization and resilience: Experiences of rural LGBTQ+ youth across multiple contexts. *Journal of Interpersonal Violence*, *37*(19-20), NP18988-NP19015.

World Health Organization. (2005). *Promoting mental health: Concepts, emerging evidence, practice.* https://www.who.int/publications/i/item/9241562943

Yang, C. (2022). Social media social comparison and identity processing styles: Perceived social pressure to be responsive and rumination as mediators. *Applied Developmental Science*, *26*(3), 504–515. https://doi.org/10.1080/10888691.2021.1894149

Yang, C., Holden, S. M., & Carter, M. D. K. (2018). Social Media Social Comparison of Ability (but not Opinion) Predicts Lower Identity Clarity: Identity Processing Style as a Mediator. *Journal of Youth and Adolescence*, *47*(10), 2114–2128. https://doi.org/10.1007/s10964-017-0801-6

Yang, C., Holden, S. M., Carter, M. D. K., & Webb, J. J. (2018). Social media social comparison and identity distress at the college transition: A dual-path model. *Journal of Adolescence*, *69*(1), 92–102. https://doi.org/10.1016/j.adolescence.2018.09.007

Yang, C., Pham, T., Ariati, J., Smith, C., & Foster, M. D. (2021). Digital Social Multitasking (DSMT), Friendship Quality, and Basic Psychological Needs Satisfaction Among Adolescents: Perceptions as Mediators. *Journal of Youth and Adolescence*, *50*(12), 2456–2471. https://doi.org/10.1007/s10964-021-01442-y

Zimmer-Gembeck, M. J., Hawes, T., & Pariz, J. (2021). A closer look at appearance and social media: Measuring activity, self-presentation, and social comparison and their associations with emotional adjustment. *Psychology of Popular Media*, *10*, 74–86. https://doi.org/10.1037/ppm0000277

Zimmer-Gembeck, M. J., Rudolph, J., Webb, H. J., Henderson, L., & Hawes, T. (2021a). Face-to-Face and Cyber-Victimization: A Longitudinal Study of Offline Appearance Anxiety and Online Appearance Preoccupation. *Journal of Youth and Adolescence*, *50*(12), 2311–2323. https://doi.org/10.1007/s10964-020-01367-y

## Table 1: Participant Information

| Participant ID | Field[a] | Country (State) | Interview Participant (N=25) | Convening Participant (N=26) |
|---|---|---|---|---|
| P1 | Communications | Belgium | x | x |
| P2 | Law | USA (WA) | x | |
| P3 | Psychology | USA (UT) | x | |
| P4 | Global Health | USA (WA) | x | x |
| P5 | Clinical Psychology | USA (NY) | x | x |
| P6 | Communications | Netherlands | x | x |
| P7 | Clinical Psychology | USA (CA) | x | |
| P8 | Sociology/Anthropology/Science & Technology Studies | USA (MA) | x | |
| P9 | Developmental Science | USA (MA) | x | x |
| P10 | Communications | UK | x | |
| P11 | Educational Sciences | Spain | x | |
| P12 | Clinical Psychology | USA (RI) | x | x |
| P13 | Computer Science | Ireland | x | x |
| P14 | Developmental Psychology | USA (MA) | x | x |
| P15 | Psychology | Australia | x | |
| P16 | Public Health | Italy | x | |
| P17 | Clinical Psychology | Germany | x | |
| P18 | Clinical Psychology | USA (OR) | x | |
| P19* | Human-Computer Interaction (HCI) | UK | x | x |
| P20 | Clinical Psychology | USA (NC) | x | x |
| P21 | Clinical Psychology | New Zealand | x | |
| P22* | Human Development/Educational Psychology | USA (OK) | x | x |
| P23 | HCI | USA (FL) | x | |
| P24 | Education/Information Science | Canada | x | x |
| P25* | Educational Psychology | UK | x | x |
| P26 | Sociology/Science & Technology Studies | USA (NY) | | x |
| P27 | Sociology | USA (CA) | | x |
| P28 | Clinical Child Psychology | USA (WA) | | x |
| P29 | Media Studies | Colombia | | x |
| P30 | Psychology | UK | | x |
| P31 | Cybersecurity | UK | | x |
| P32* | Clinical Psychology | USA (PA) | | x |
| P33* | School Psychology | USA (WA) | | x |
| P34 | HCI | USA (WA) | | x |
| P35 | HCI | UK | | x |
| P36 | HCI | USA (WA) | | x |
| P37 | HCI | USA (WA) | | x |
| P38* | Human Development & Education/HCI | USA (WA) | | x |

*Co-author
[a]Self-identified field/s. During the convening, participants observed that several of these fields overlap.

**Table 2. Summary of notable developments, persistent gaps, and paths forward for each of the three focal topics.**

|  | Measurement | Representation | Intervention |
|---|---|---|---|
| **Notable Developments (RQ1)** | - shift away from "screen time" as dominant measure of tech use<br>- greater attention to the *quality* of youth's tech experiences<br>- increase in person-specific analyses<br>- increase in longitudinal studies | - increased representation of diverse adolescents (e.g., diversity across culture, race, gender, sexuality)<br>- greater attention to the intersections of youth's marginalized identities<br>- inclusion of youth's voices in the research process | - development of school-based curricula targeting digital well-being<br>- technology-based interventions providing in-the-moment support<br>- leveraging technology as a delivery mechanism for mental health support |
| **Persistent Gaps & Tensions (RQ2)** | - persistence of cross-sectional studies that use screen time as primary measure<br>- methodological disagreement about person-specific analyses<br>- lack of clarity and uniformity in definition and measurement of social media use | - paucity of cross-cultural research<br>- scarcity of research focused on youth living in the Global South<br>- danger in replicating existing stereotypes and biases through research | - few interventions (e.g., digital self-control tools) developed specifically for teens<br>- few commercial interventions target diverse populations (e.g., Spanish-speaking youth)<br>- challenge of bringing research-based interventions to a broader audience (implementation) |
| **Paths Forward (RQ3)** | - continue/increase cross-disciplinary collaborations<br>- greater transparency and rigor in research methods, e.g., open science model<br>- increased focus on *mechanisms* that lead to positive or negative outcomes | - continue to expand the diversity of youth perspectives represented<br>- greater attention to researcher positionality<br>- need for greater researcher diversity<br>- expand participation of youth directly in research | - shift perspective beyond a Western therapy model<br>- include people from diverse backgrounds in intervention development<br>- Study real-world impact of interventions |